\documentclass[reprint,pre,superscriptaddress,amsmath,amssymb,aps,
]{revtex4-2}

\usepackage{bm}%
\usepackage{wasysym}
\usepackage{amssymb}
\usepackage{bbold}
\usepackage[colorlinks=true,linkcolor=blue]{hyperref}%
\expandafter\ifx\csname package@font\endcsname\relax\else
 \expandafter\expandafter
 \expandafter\usepackage
 \expandafter\expandafter
 \expandafter{\csname package@font\endcsname}%
 \fi
\usepackage{graphicx}
\usepackage{dcolumn}

\begin{document}

\title{Multifractal and Ergodic Properties of Conductance Fluctuations under Strong Disorder}

\author{Marcos A. A. de Sousa}
\affiliation{Departamento de F\'{\i}sica, Universidade Federal Rural de Pernambuco, Recife - PE, 52171-900, Brazil}

\author{Heitor R. Publio}
\affiliation{Departamento de F\'{\i}sica, Universidade Federal Rural de Pernambuco, Recife - PE, 52171-900, Brazil}

\author{Henrique A. de Lima}
\affiliation{International Center of Physics, Institute of Physics, University of Brasilia, 70910-900 Brasilia, Federal District, Brazil}

\author{Adauto J. F. de Souza}
\affiliation{Departamento de F\'{\i}sica, Universidade Federal Rural de Pernambuco, Recife - PE, 52171-900, Brazil}

\author{Fernando A. Oliveira}
\email{faooliveira@gmail.com}
\affiliation{Departamento de F\'{\i}sica, Universidade Federal Rural de Pernambuco, Recife - PE, 52171-900, Brazil}
\affiliation{International Center of Physics, Institute of Physics, University of Brasilia, 70910-900 Brasilia, Federal District, Brazil}

\author{Anderson L. R. Barbosa}
\email{anderson.barbosa@ufrpe.br}
\affiliation{Departamento de F\'{\i}sica, Universidade Federal Rural de Pernambuco, Recife - PE, 52171-900, Brazil}

\begin{abstract}
Understanding the stochastic properties of conductance fluctuations in disordered mesoscopic systems is fundamental to quantum transport. In this work, we investigate the multifractal and ergodic properties of the fictitious time series of conductance in two-dimensional tight-binding models under varying Anderson disorder. Using standard multifractal analysis, we show that conductance fluctuations exhibit a transition from non-ergodic to ergodic behavior as the disorder strength increases, as evidenced by the decay of the conductance correlation function. Remarkably, multifractality persists in both regimes; however, it becomes insensitive to shuffling in the strong-disorder (ergodic) regime, suggesting that distributional effects dominate temporal organization. On the contrary, in the weakly disordered (non-ergodic) regime, long-range correlations play a significant role. These findings are robust against changes in lead geometry (asymmetric vs. symmetric). Our results provide new insights into the interplay between ergodicity, multifractality, and rare events in disordered quantum transport.

\end{abstract}

\maketitle

\section{Introduction} 

Conductance fluctuations are among the most remarkable phenomena in mesoscopic physics, arising from quantum interference effects resulting from multiple wave scatterings within a sample \cite{Beenakker_1997,RevModPhys.89.015005,PhysRevLett.55.1622,xv5t-qvcm,PhysRevB.88.245133,Amin_2018,PhysRevE.104.054129}. Conductance fluctuations have been observed across a wide range of phase-coherent electron transport systems, from diffusive nanowires—where soft disorder scattering induces Anderson localization when the sample length exceeds the localization length—to ballistic chaotic billiards, where elastic scattering occurs at the boundaries \cite{PhysRevLett.77.3885,PhysRevB.98.155407,PhysRevB.102.115105,PhysRevLett.80.1948,PhysRevB.58.11107,PhysRevLett.78.1952,PhysRevB.54.10841}.

Conductance fluctuations can be interpreted as arising from stochastic systems \cite{Amin_2018,PhysRevE.104.054129,PhysRevLett.128.236803,PhysRevB.111.L081405,PhysRevLett.131.076301}. In this context, conductance is treated as a function of a control parameter, such as Fermi energy, gate voltage, or magnetic field, effectively creating a {\it fictitious time series} \cite{PhysRevLett.70.4122,BEENAKKER199461,PhysRevLett.79.913}. As demonstrated in Refs. \cite{Amin_2018,PhysRevE.104.054129}, the conductance fictitious time series through a nanowire exhibits multifractal features when influenced by weak Anderson disorder. This multifractality arises from correlations induced by modulations in the density of states resulting from variations in the control parameters \cite{PhysRevE.104.054129}. However, in the strong Anderson disorder regime, the stochastic properties of the conductance fictitious time series are missing. 

In the strongly Anderson disorder regime, quantum transport is highly inhomogeneous, following directed paths—an analogy with the directed polymer problem, one of the simplest statistical physics models where disorder plays a nontrivial role \cite{PhysRevB.93.054201,Prior2009,PhysRevB.91.155413,Derrida1988,PhysRevLett.122.030401}. More recently, it was revealed that quantum transport under strong Anderson disorder exhibits glassy properties in a two-dimensional device at zero temperature, which includes pinning, avalanches, and chaos \cite{PhysRevLett.122.030401}.

On the other hand, numerical investigations of level and eigenfunction statistics on random regular graphs under strong disorder revealed evidence for an ergodic transition from extended non-ergodic (multifractal) to extended ergodic states, characterized by a jump in fractal dimensions rather than a smooth crossover~\cite{PhysRevLett.113.046806,Altshuler_2016,Facoetti_2016,Kravtsov_2015}. Recent work has demonstrated the existence of this non-ergodic extended phase and its transition to the ergodic extended phase~\cite{KRAVTSOV2018148}. 

Khinchin proved a long time ago that if a time series exhibits a time correlation function \( C(\delta t) \) that satisfies the mixing condition \( C(\delta t \rightarrow \infty) = 0 \), then ergodicity is achieved in any physical system \cite{Khinchin49,Lee07}, where \( \delta t \) represents the time step. Specifically, a finite time correlation length indicates a non-ergodic system, while an approach toward zero time correlation length signifies an ergodic system. Lapas et al. \cite{Lapas08,Lapas_2007} demonstrated that this principle also holds in cases with strong correlation and memory effects, for review see~\cite{GOMESFILHO25}.

In this context, we will use a fictitious time series of conductance to illustrate that conductance fluctuations share a key property with stochastic systems: they can be classified as either ergodic or non-ergodic. In particular, we show that there is a transition from non-ergodicity to ergodicity as a function of Anderson disorder. In the presence of weak disorder, the transport dynamics are non-ergodic; conversely, with strong disorder, they become ergodic. Furthermore, the fictitious time series in both regimes exhibit multifractality, although the origins of their multifractality differ.
Multifractality is consistent with multiplicative temporal organization in the weakly disordered regime \cite{PhysRevE.107.034139,KELTYSTEPHEN2024129573,PhysRevE.109.064212}, whereas its origin shifts to distribution-dominated fluctuations in the strongly disordered regime.
This disorder-dependent phenomenon cannot be explained as a mere artifact, such as a finite-sized time series \cite{PhysRevE.107.034139}, and may indeed signal a structural change in the underlying dynamics.

This work is organized as follows: Section 2 presents our methodology, including our microscopic model, the temporal evolution of conductance, and the correlation function. We also describe the procedure for our multifractal analysis. Section 3 details the results and discussions, and we conclude in Section 4.

\section{Methodology}

In this section, we present the tight-binding Hamiltonian model and provide a brief introduction to Multifractal Detrended Fluctuation Analysis (MF-DFA) \cite{KANTELHARDT200287}, which we employ to derive our results, as shown in the next section.

\begin{figure}
    \centering
    \includegraphics[width=1.0\linewidth]{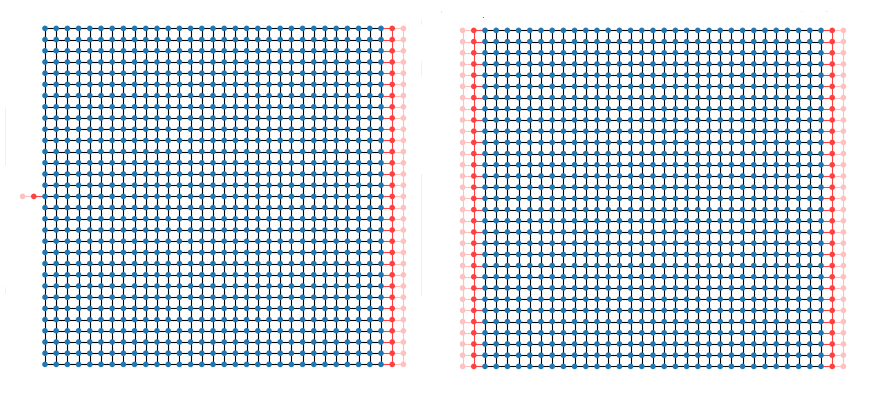}
    \caption{Mesoscopic two-dimensional sample (in blue) connected to asymmetric (on the left) and symmetric (on the right) leads (in red) while being subjected to strong disorder. Both samples have dimensions of \( 61a \times 61a\), where \(a\) is the lattice constant.}
    \label{fig:sample}
\end{figure}

\subsection{Microscopic model}

We explore electronic transport through a mesoscopic two-dimensional (2D) sample with strong disorder, connected to asymmetric and symmetric leads, as illustrated in Figure~\ref{fig:sample}. The scattering matrix that describes the electronic transport through the 2D sample is given by
\[
\bf{S} = 
\begin{bmatrix}
{\bf{r}} & {\bf{t}'} \\
{\bf{t}} & {\bf{r}'}
\end{bmatrix},
\]
where \( \bf{t} \) and \( \bf{t}' \) represent the transmission blocks, while \( \bf{r} \) and \( \bf{r}' \) denote the reflection blocks. We can interpret \(\bf{t}'\) as the transmission block from right to left leads and \(\bf{t}\) from left to right leads, while \(\bf{r}'\) is the reflection block from right to right lead and \(\bf{r}\) left to left lead. The conductance can be calculated using the Landauer-Büttiker relation
\[
G = \frac{2e^2}{h}\text{Tr}[\bf{t} {\bf{t}}^\dagger], \label{T}
\]
which is valid at the linear response regime and low temperatures. We performed numerical calculations of the conductance using the KWANT software \cite{Groth_2014}, which employs Green's-function methods within the tight-binding framework.
The tight-binding Hamiltonian for graphene is defined as follows
\[
\hat{H} = t \sum_{\langle i, j \rangle} c_{i}^\dagger c_{j} + \sum_{i} \epsilon_i c_{i}^\dagger c_{i}, \label{H}
\]
where the indices \( i \) and \( j \) span all lattice sites, and \( \langle i, j \rangle \) indicates first nearest neighbors. The first term in \( \hat{H} \) represents the typical electron hopping \( t \) between lattice sites, with \( c_{i} \) and \( c_{i}^{\dagger} \) being the annihilation and creation operators, respectively. The disorder is realized by an on-site electrostatic potential \(\epsilon_i\), which varies randomly from site to site according to a uniform distribution in the interval \( \left( -U/2, U/2\right) \), where \( U \) is the disorder width.

\begin{figure}
    \centering
    \includegraphics[width=1.0\linewidth]{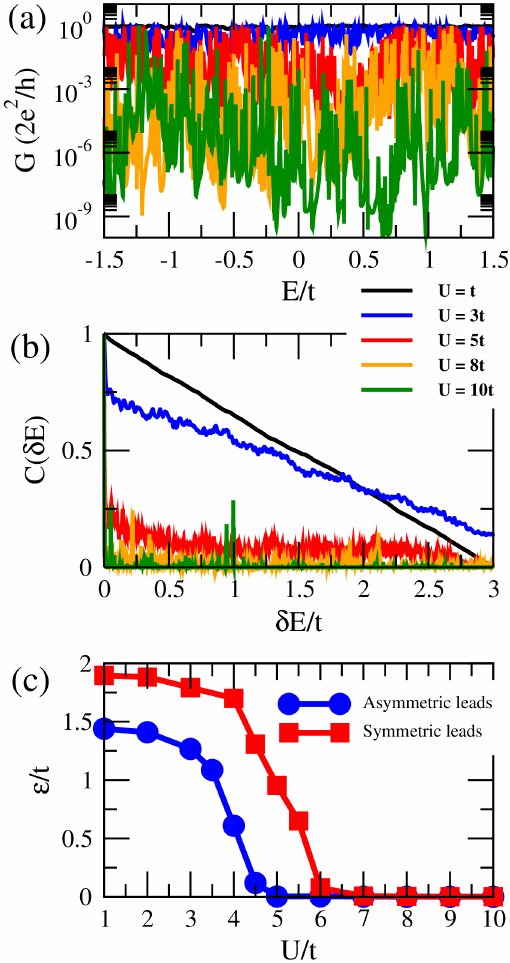}
    \caption{Panel (a) shows the conductance as a function of the Fermi energy \(E\) from a 2D sample with asymmetric leads for different values of disorder \(U\), as illustrated in Figure~\ref{fig:sample}. Panel (b) shows the conductance correlation \(C(\delta E)\) calculated from the conductance fictitious time series as illustrated in panel (a). Panel (c) shows the fictitious correlation time \(\epsilon\) as a function of disorder \(U\) for a 2D sample connected to asymmetric (circle symbols) and symmetric (square symbols) leads.}
    \label{fig:energy}
\end{figure}

The upper panel in Figure~\ref{fig:energy} shows the conductance as a function of the Fermi energy \(E\) from a 2D sample with asymmetric leads for different values of disorder \(U\). At weak disorder, interference effects and mesoscopic conductance fluctuations dominate electron transport; however, at strong disorder, rare events take precedence, as we will demonstrate.

\subsection{Multifractal Analysis}
\label{MultifractalAnalysis}

We briefly introduce the Multifractal Detrended Fluctuation Analysis (MF-DFA) \cite{KANTELHARDT200287}, applied to the time series of physical observables. For concreteness, we focus on the conductance {\it fictitious time series}, where the Fermi energy is interpreted as a {\it fictitious time}.

Let $g_t$ denote the dimensionless conductance \(g = G/\left(2e^2/h\right)\) in the $t$-th time step, $t=1,2,\dots,N$, where $N$ is the total number of time steps. We first define the profile of $\{g_t\}$ as follows
\begin{equation}
    \tilde{g}(i) = \sum_{t=1}^i \left(g_t - \langle g \rangle\right),
\end{equation}
where
\begin{equation}
    \langle g \rangle = \frac{1}{N} \sum_{t=1}^{N} g_t
\end{equation}
is the average dimensionless conductance time series.
The profile $\{\tilde{g}(i)\}$ is divided into $N_s = \textrm{int}(N/s)$ non-overlapping segments of size $s$, denoted $\{\tilde{g}_j(i)\}$, $j=1,\dots,N_s$. In each segment, the local trend is removed by fitting a linear function $f_j(i)$, and the variance is calculated as
\begin{equation}
    F_s^2(j) = \frac{1}{s} \sum_{i=1}^s \big\{\tilde{g}[(j-1)s+i] - f_j(i)\big\}^2.
    \label{Fs}
\end{equation}
Next, the $q$-th order fluctuation function is defined as
\begin{equation}
    F_q(s) = \left( \frac{1}{2N_s} \sum_{j=1}^{2N_s} \big[F_s^2(j)\big]^{q/2} \right)^{1/q},
    \label{Fp}
\end{equation}
where $q$ takes several real values: $\pm0.2$, $\pm0.4$, $\pm0.6$, $\pm0.8$, $\pm1.0$, $\pm2.0$, $\pm3.0$, $\pm4.0$, and $\pm5.0$.  
The sum in Eq. \ref{Fp} is done from both the beginning and the end of the time series, which leads to $2N_s$ segments in total \cite{Zhao2016Multifractality}. The scaling behavior of $F_q(s)$ with segment size $s$ is characterized by the generalized Hurst exponent $h(q)$:
\begin{equation}
    F_p(s) \sim s^{h(q)}.
    \label{hq}
\end{equation}
If $h(q)$ depends on $q$, the series is multifractal, while if $h(q)$ is independent of $q$, it is monofractal.
The multifractal spectrum is obtained via~\cite{Halsey86}
\begin{equation}
    \tau(q) = q h(q) - 1,
\end{equation}
and the singularity spectrum $f(\alpha)$ is defined through a Legendre transform:
\begin{equation}
    f(\alpha) = q \alpha - \tau(q), \quad \alpha = \frac{d \tau}{dq}.
    \label{fa}
\end{equation}
Multifractal series are characterized by a broad spectrum $f(\alpha)$, while monofractal series produce a narrow spectrum. The strength of multifractality can be quantified by the width $\Delta \alpha = \alpha_{\max} - \alpha_{\min}$; As $\Delta \alpha \to 0$, the series approaches monofractal behavior.

\begin{figure*}
    \centering
    \includegraphics[width=1.0\linewidth]{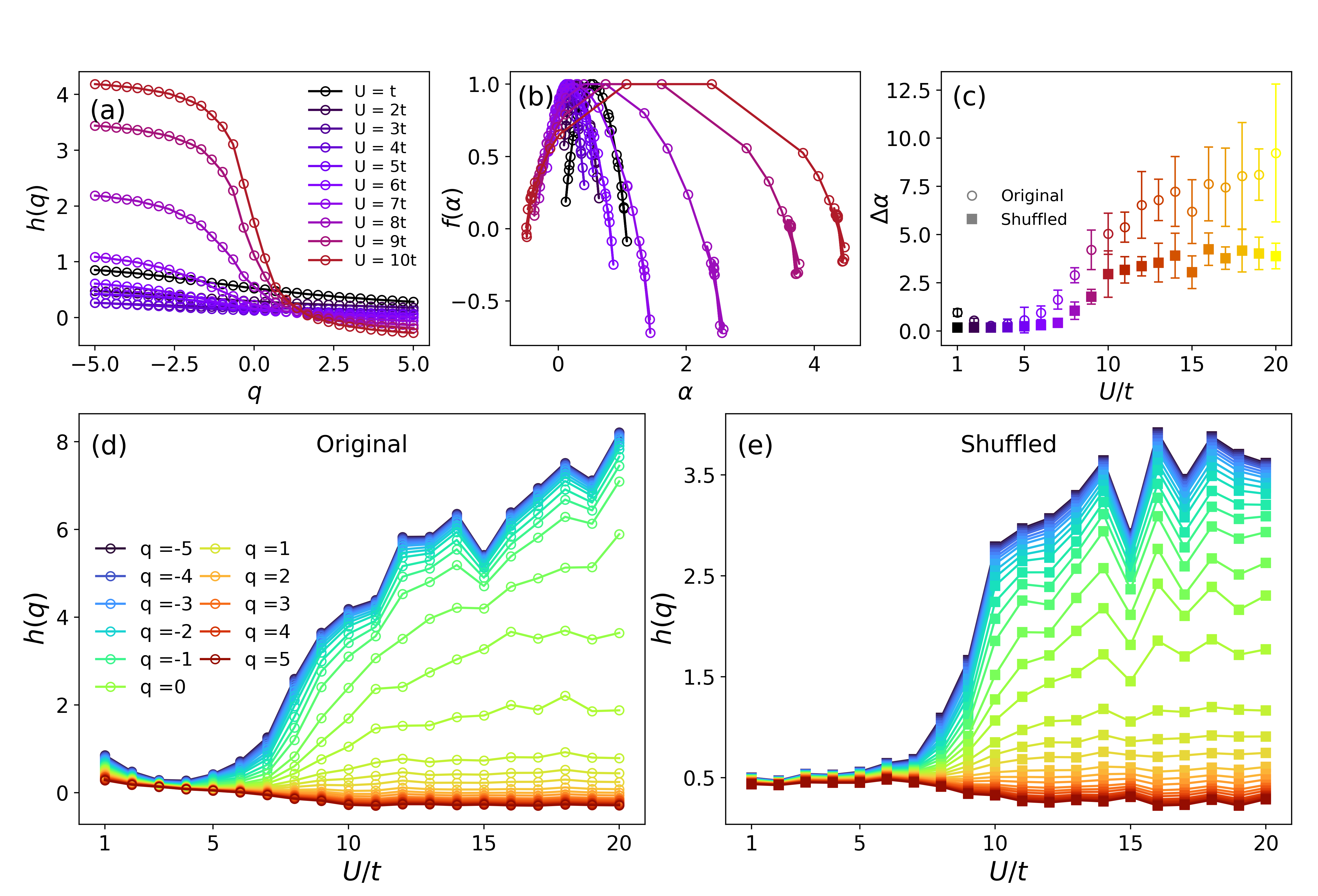}
    \caption{Panel (a) presents the generalized Hurst exponents, \(h(q)\), and panel (b) shows the multifractal singularity spectra, \(f(\alpha)\), for the conductance time series of a 2D sample with asymmetric leads, as illustrated in Figure~\ref{fig:energy}, at various disorder values \(U\). Panel (c) plots the width \( \Delta \alpha = \alpha_{\max} - \alpha_{\min}\), derived from panel (b), as a function of \(U\) for original (circle symbols) and shuffled (square symbols) time series. Panel (d) shows \(h(q)\) versus \(U\), and panel (e) shows the Hurst exponent from shuffled time series as a function of \(U\).}
    \label{fig:1dmulti}
\end{figure*}

\begin{figure}
    \centering
    \includegraphics[width=1.0\linewidth]{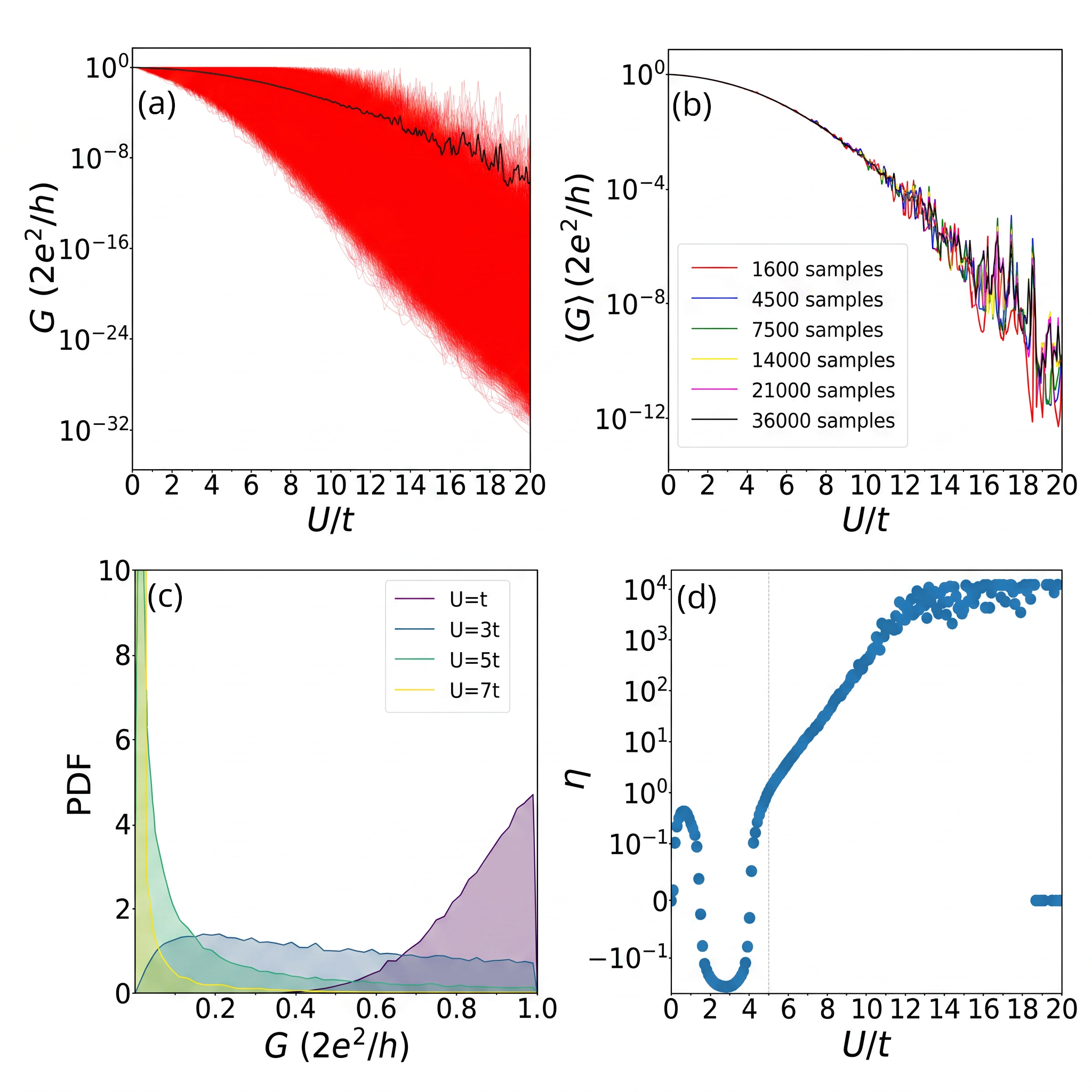}
    \caption{Panel (a) illustrates the conductance as a function of disorder \(U\) for 36,000 different 2D samples with asymmetric leads, represented by the red dotted lines. The black line depicts the average conductance across these 36,000 samples. Panel (b) presents the average conductance calculated from varying numbers of samples. Panel (c) shows the probability distribution function (PDF) of conductance for different values of \(U\). Panel (d) displays the non-Gaussian term \(\eta\), as defined in Eq.~\ref{nG}, as a function of \(U\).   }
    \label{fig:distribution}
\end{figure}

\section{Results and Discusion}

From the conductance fictitious time series presented in Figure~\ref{fig:energy}(a), we can calculate the  fictitious conductance correlation defined by
\[C(\delta E) = \text{corr}[G(E+\delta E), G(E)]/\left(2e^2/h\right)^2,\] 
where \(\delta E=10^{-3}t\). This correlation is shown in Figure~\ref{fig:energy}(b) for various disorder values. As expected, the fictitious correlation time \(\epsilon\) decreases as the disorder \(U\) increases. We can determine \(\epsilon\) for each disorder value, and the relationship between fictitious correlation time and disorder \(U\) is shown in Figure~\ref{fig:energy}(c). For a 2D sample with asymmetric leads (denoted by circular symbols), the \(\epsilon\) exhibits a distinct transition from non-ergodic to ergodic behavior~\cite{Lapas_2007,Costa06}, indicating a shift in the mechanisms underlying fluctuations in conductance time series. The rapid decay of the fictitious correlation function suggests a crossover toward effectively ergodic transport dynamics.
Notice that in a fictitious time-dependent correlation function $C(\delta E)$,  the limit $C(\delta E \rightarrow \infty) \rightarrow 0$ is named mixing. This guarantees ergodicity in the stochastic process \cite{Lapas08}. Additionally, a 2D sample with symmetric leads (marked by square symbols) exhibits a similar trend. These findings are consistent with the results of Ref. \cite{PhysRevLett.122.030401}.

To fully understand the consequences of the transition from non-ergodic to ergodic behavior on the mechanisms behind fluctuations in conductance time series, we performed an extensive multifractal analysis of these series. For each disorder value \(U\), we calculated ten distinct conductance time series from ten distinct samples. All results we present are averages of ten conductance time series.

\subsection{2D sample with asymmetric leads}

Figure~\ref{fig:1dmulti}(a) shows the Hurst exponent average \(h(q)\) as a function of \(q\) calculated from Eq.~\ref{hq}, for various disorder values \(U\). At \(U=t\), the Hurst exponent demonstrates a significant dependence on \(q\), indicating clear multifractality in the conductance time series. As the disorder increases, the energy correlation length decreases, leading to \(h(q)\) becoming independent of \(q\). This trend indicates that the conductance time series tend to be monofractal. These results agree with Refs. \cite{Amin_2018,PhysRevE.104.054129}.
In contrast, for \(U>5t\), where \(\epsilon \rightarrow 0\), \(h(q)\) begins to change with \(q\), signifying a return to multifractality. This behavior is more evident when plotting the Hurst exponent as a function of disorder \(U\), as shown in Figure~\ref{fig:1dmulti}(d). 

To further support the results of the Hurst exponent, we also analyze the corresponding singularity spectrum, obtained through a Legendre transformation as described in Eq.~\ref{fa}. Figure~\ref{fig:1dmulti}(b) displays the function \( f(\alpha) \) for various values of \(U\). The broad and pronounced width, \( \Delta \alpha \), of this spectrum provides strong evidence of multifractality in the presence of strong disorder. Next, in Figure~\ref{fig:1dmulti}(c), we plot \( \Delta \alpha \) as a function of \(U\) for both original and surrogate conductance time series. Initially, \( \Delta \alpha \) decreases as \(U\) increases, but then, for \(U>5t\), \( \Delta \alpha \) begins to rise with increasing disorder, supporting our earlier findings. Additionally, for \(U \approx 13t\), \( \Delta \alpha \) becomes largely independent of disorder. Notably, multifractality persists in the strongly disordered region for surrogate series, whereas in the weakly disordered regime it remains insensitive to shuffling.


The main source of intrinsic multifractality in a time series is typically long-range correlations \cite{KANTELHARDT200287,PhysRevLett.128.236803}, which should be destroyed by shuffling \cite{PhysRevE.107.034139}. Apparent multifractality may persist after shuffling due to finite-size effects, heavy-tail-induced spectrum broadening, or nonlinear correlations \cite{PhysRevE.107.034139,KELTYSTEPHEN2024129573}. Therefore, fat-tailed distributions of the series values, where multifractality is not removed through shuffling, may be due to non-Gaussian sampling effects. To investigate the origin of multifractality in the fluctuations of conductance time series, we computed the Hurst exponent for the corresponding shuffled conductance time series, as shown in Figure~\ref{fig:1dmulti}(e). 
For \(U < 5t\), we found \(h(q) \approx 0.5\), suggesting that the shuffled series are monofractal and uncorrelated. This suggests that the multifractality in the conductance time series is mainly due to correlations induced by variations in the Fermi surface. In this regime, interference effects and mesoscopic conductance fluctuations are dominant. 

In contrast, for \(U > 5t\), the Hurst exponent exhibits a significant dependence on \(q\), indicating clear multifractality in the shuffled conductance time series. This finding is supported by Figure~\ref{fig:1dmulti}(c), where we plot \( \Delta \alpha \) as a function of \(U\) for shuffled time series. We observe that \( \Delta \alpha \) is approximately null for \(U < 5t\) and increases for \(U > 5t\).  The persistence of multifractal scaling after shuffling suggests that the observed spectrum is dominated by non-Gaussian distributional effects rather than temporal correlations.
This implies that in non-ergodic transport dynamics, multifractal conductance fluctuations are induced by correlations. Conversely, in ergodic transport dynamics, these fluctuations arise from the non-Gaussianity of the conductance fluctuations, which cannot be eliminated by shuffling the conductance time series. Notice that, surrogate analysis is insufficient to separate nonlinear correlations from probability distribution function effects. To the best of our knowledge, such a disorder-driven crossover in the sensitivity of multifractal scaling to reshuffling has not been reported in conductance fluctuations.


To confirm that fat-tailed distributions of conductance are prevalent in the presence of strong disorder, we conducted numerical calculations of conductance as a function of \(U\) across 36,000 different samples, as illustrated in Figure~\ref{fig:distribution}(a). The red dashed lines represent the 36,000 samples, while the solid black line indicates the average conductance. It is important to note that the maximum conductance is limited by \(G/(2e^2/h) = 1\), since the 2D sample with asymmetric leads has only one propagating wave channel in the left lead (see Figure~\ref{fig:sample}).

In Figure~\ref{fig:distribution}(b), we display the average conductance for varying numbers of samples. In the presence of strong disorder, fluctuations in conductance do not diminish because rare events dominate the average, highlighting the presence of fat-tailed conductance distributions. Consequently, we present the conductance probability distribution function (PDF) for different values of disorder in Figure~\ref{fig:distribution}(c). For \(U=t\), the PDF peaks at \(G/(2e^2/h) = 1\) and does not exhibit a fat tail. At \(U=3t\), the PDF is spread between 0 and 1 and lacks fat tails. In contrast, for \(U=5t\) and \(U=7t\), the PDFs peak at 0 and demonstrate the presence of fat tails, clearly indicating the dominance of fat-tailed distributions of conductance under strong disorder.

We can quantify how fat-tails affect the conductance PDF by calculating the unidimensional non-Gaussian indicator \cite{Lapas_2007,Costa06}, defined by the equation  
\begin{equation}
\eta  = \frac{\langle G^4 \rangle}{3\langle G^2 \rangle^2}-1.\label{nG}    
\end{equation}
The closer \(|\eta|\) is to zero, the more it resembles a Gaussian distribution; conversely, the larger \(|\eta|\) becomes, the less it resembles the Gaussian distribution.
Figure~\ref{fig:distribution}(d) illustrates the behavior of \(\eta\) as a function of \(U\). For \(U < 5t\), the deviation of the PDF from a Gaussian shape is small, with \(-1 < \eta < 1\). In this regime, the conductance fluctuations are non-ergodic and predominantly influenced by interference effects. In contrast, for \(U > 5t\), \(\eta\) increases exponentially with \(U\). Notably, around \(U \approx 13t\), \(\eta\) begins to show independence from disorder, which aligns with the results of the multifractal analysis presented in Figure~\ref{fig:1dmulti}(c). Therefore, in this regime, conductance fluctuations are ergodic and exhibit fat-tailed distributions, with rare events playing a significant role in the conductance dynamics.

\begin{figure*}
    \centering
    \includegraphics[width=0.9\linewidth]{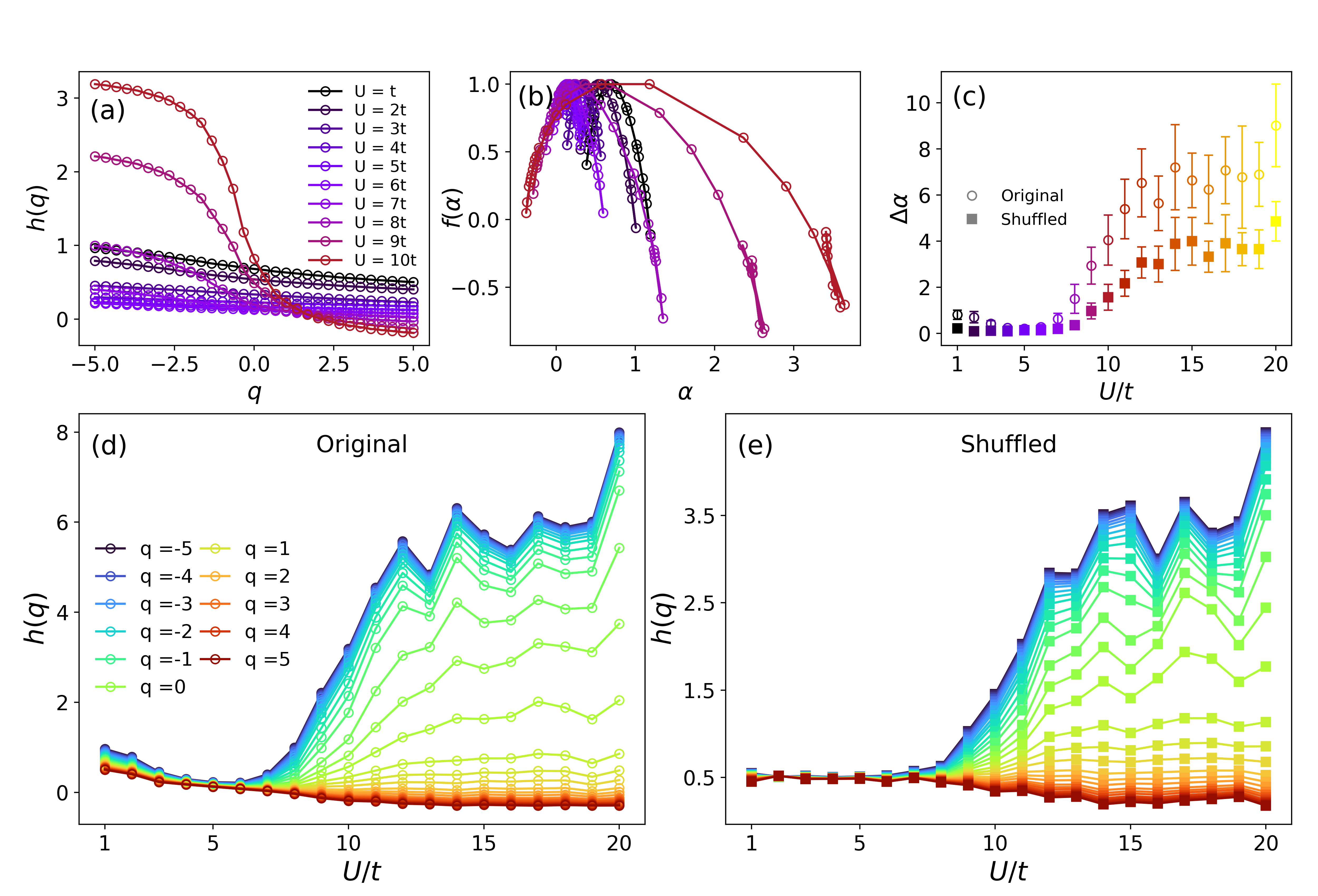}
    \caption{Panel (a) presents the generalized Hurst exponents, \(h(q)\), while panel (b) shows the multifractal singularity spectra, \(f(\alpha)\), associated with the conductance time series of a 2D sample with symmetric leads. These panels represent different values of disorder \(U\). Panel (c) displays the width \(\Delta \alpha = \alpha_{\max} - \alpha_{\min}\), derived from panel (b), as a function of \(U\) for original (circle symbols) and shuffle (square symbols) time series. In panel (d), the Hurst exponents, \(h(q)\), are plotted as a function of \(U\). Panel (e) shows the Hurst exponent calculated from shuffled conductance time series as a function of \(U\).}
    \label{fig:2dmulti}
\end{figure*}

\subsection{2D sample with symmetric leads}

To confirm the general idea of our analysis, we also conducted a systematic multifractal analysis of the conductance time series for a 2D sample with symmetric leads, as illustrated in Figure~\ref{fig:sample}. 

Figure~\ref{fig:2dmulti}(a) presents the average Hurst exponent \(h(q)\) as a function of \(q\) for various disorder levels \(U\). At \(U = t\), the Hurst exponent exhibits a significant dependence on \(q\), indicating clear multifractality in the conductance time series. However, as disorder increases, the energy correlation length decreases, resulting in \(h(q)\) becoming independent of \(q\). This trend implies that the conductance time series tends towards monofractality.
In contrast, for \(U > 7t\), where \(\epsilon \rightarrow 0\), \(h(q)\) begins to change with \(q\), signaling a return to multifractality. This behavior becomes more evident when we plot the Hurst exponent as a function of disorder \(U\), as shown in Figure~\ref{fig:2dmulti}(d). 

To further support our findings on the Hurst exponent, we analyzed the corresponding singularity spectrum. Figure~\ref{fig:2dmulti}(b) shows \(f(\alpha)\) for different values of \(U\). The broad shape and significant width of the spectrum, \(\Delta \alpha\), provide strong evidence of multifractality in the presence of strong disorder. In Figure~\ref{fig:2dmulti}(c), we plot \(\Delta \alpha\) as a function of \(U\). Initially, we observe a decrease in \(\Delta \alpha\) with increasing \(U\), followed by an increase in \(\Delta \alpha\) for \(U > 7t\), which confirms our findings. Additionally, for \(U \approx 13t\), \(\Delta \alpha\) starts to become independent of disorder.

To investigate the origin of multifractality in the fluctuations of conductance time series, we computed the Hurst exponent for the corresponding shuffled conductance time series, as shown in Figure~\ref{fig:2dmulti}(e). For \(U < 7t\), we found that \(h(q) \approx 0.5\), indicating that the shuffled series is monofractal and uncorrelated. This suggests that the multifractality in the conductance time series is primarily due to correlations induced by variations in the Fermi surface. In this regime, interference effects and mesoscopic conductance fluctuations are predominant.

In contrast, for \(U > 7t\), the Hurst exponent exhibits a significant dependence on \(q\), which highlights clear multifractality in the shuffled conductance time series. This conclusion is further supported by Figure~\ref{fig:2dmulti}(c), which shows the variation of \( \Delta \alpha \) as a function of \(U\) for shuffled time series. Specifically, we observe that \( \Delta \alpha \) is approximately zero for \(U < 7t\), but increases for \(U > 7t\). As before, this multifractality can be attributed to the non-Gaussianity of the conductance fluctuations. Therefore, in non-ergodic transport dynamics, multifractal conductance fluctuations result from correlations. In contrast, in ergodic transport dynamics, they are strongly influenced by non-Gaussian distributional effects rather than temporal correlations. In the latter case, we note that shuffling alone is insufficient to eliminate multifractality in the conductance time series.


These analyses indicate that for a two-dimensional sample, connecting to either asymmetric or symmetric leads produces similar physical outcomes. Both types of connections exhibit a transition from non-ergodic to ergodic conductance dynamics as disorder increases. In the non-ergodic regime, conductance dynamics are primarily influenced by interference effects. In contrast, in the ergodic regime, the dynamics become strongly influenced by non-Gaussian distributional effects.

\section{Conclusions}
In this work, we systematically analyzed the multifractal and ergodic properties of conductance fluctuations in a two-dimensional disordered mesoscopic system, modeled via a tight-binding Hamiltonian with on-site disorder. By treating conductance as a function of Fermi energy—i.e., as a fictitious time series—we applied MF-DFA to characterize the generalized Hurst exponent and singularity spectrum across a wide range of disorder strengths \(U\).

Our main findings are threefold. First, we observed a clear transition from non-ergodic to ergodic transport dynamics as \(U\) increased, as evidenced by the vanishing of the fictitious correlation time \(\epsilon\). This transition is independent of lead geometry and occurs in both asymmetric and symmetric lead configurations. Second, multifractality is present in both the weak- and strong-disorder regimes, but its microscopic origin differs qualitatively. In the non-ergodic regime (\(U<5t\) for asymmetric leads, \(U<7t\) for symmetric leads), multifractality is primarily due to long-range correlations, as confirmed by shuffling the time series. In the ergodic regime (\(
U>5t/7t\)), multifractality survives. The persistence of multifractal scaling after shuffling suggests that the observed spectrum is dominated by non-Gaussian distributional effects rather than temporal correlations. Third, we linked the emergence of fat tails to rare-event-dominated transport, quantified by the non-Gaussian indicator \(\eta\), which increases exponentially with disorder before saturating near \(U\approx13t\).
In summary, we have shown that the sensitivity of multifractality to temporal reshuffling changes across the ergodic crossover.

These results demonstrate that conductance fluctuations in disordered systems offer a rich platform where ergodicity breaking and multifractality are not merely coexistent but can be driven by distinct physical mechanisms. Our work also suggests that the ergodic transition in conductance time series may be a generic feature of Anderson localization in finite dimensions, with potential analogs in other disordered and glassy systems. 

In Ising systems \cite{Lima24,Lima25}, the connection between ergodicity, fractality, and critical exponents is more defined, although it is not yet complete even in two dimensions. Furthermore, conformal invariance \cite{Polyakov:1970xd,Polyakov:1984yq} was proven exactly for $2D$, which earned Smirnov the Fields Medal~\cite{Smirnov10,kemppainen2019}. We hoped that this could also be done for disordered Anderson systems and that this work may stimulate efforts in that direction.

\section*{Acknowledgements}

The authors acknowledge the Coordenação de Aperfeiçoamento de Pessoal de Nível Superior - CAPES.
FAO and HRP acknowledge financial support from Fundação de Amparo à Ciência e Tecnologia de Pernambuco - FACEPE (Grants APV-0064-1.05/25 and IBPG-0092-1.04/25).
ALRB and FAO acknowledge financial support from CNPq (Grants 302502/2025-4, 303119/2022-5, and 406836/2022-1 INCT of Spintronics and Advanced Magnetic Nanostructures - SpinNanoMag).

\bibliography{ref.bib}


\end{document}